\documentclass[letterpaper,twocolumn,10pt]{article}
\usepackage{usenix2019_v3}

\usepackage{tikz}
\usepackage{amsmath}
\usepackage{multirow}
\usepackage{listings}
\usepackage{lipsum} 

\usepackage{filecontents}

\begin{document}

\date{}

\title{\Large \bf Advanced Real-Time Fraud Detection Using RAG-Based LLMs}

\author{
{\rm Gurjot Singh}\\
University of Waterloo, Canada
\and
{\rm Prabhjot Singh}\\
University of Waterloo, Canada
\and
{\rm Maninder Singh}\\
Thapar Institute of Engineering and Technology, India
} 

\maketitle

\begin{abstract}
Artificial Intelligence (AI) has become a double-edged sword in modern society, being both a boon and a bane. While it empowers individuals, it also enables malicious actors to perpetrate scams such as fraudulent phone calls and user impersonations. This growing threat necessitates a robust system to protect individuals. In this paper, we introduce a novel real-time fraud detection mechanism using Retrieval Augmented Generation (RAG) technology to address this challenge on two fronts. First, our system incorporates a continuously updating policy-checking feature that transcribes phone calls in real-time and uses RAG-based models to verify that the caller is not soliciting private information, thus ensuring transparency and the authenticity of the conversation. Second, we implement a real-time user impersonation check with a two-step verification process to confirm the caller’s identity, ensuring accountability. A key innovation of our system is the ability to update policies without retraining the entire model, enhancing its adaptability. We validated our RAG-based approach using synthetic call recordings, achieving an accuracy of 97.98\% and an F1-score of 97.44\% with 100 calls, outperforming state-of-the-art methods. This robust and flexible fraud detection system is well-suited for real-world deployment.

\end{abstract}

\section{Introduction}
The issue of fraudulent activities through phone calls has grown in importance and has caused large financial losses for both individuals and organizations globally. According to a study done in 2021 by the Communications Fraud Control Association (CFCA-2021) \cite{communications2021fraud}, fraud in the telecom industry costs billions of dollars every year. In comparison to 2019 there has been a 28\% increase in fraud, or roughly USD 11.6 billion. These losses impact not only the economy as a whole but also personal security since they undermine confidence in reliable channels of communication.
One of the primary reasons for these scams' success is a lack of public awareness. Many people are unaware of the sophisticated techniques used by fraudsters and the kinds of questions they may pose. This lack of awareness makes it easier for attackers to exploit their victims. Impersonating bank representatives, government officials, or technical support agents \cite{sahin2017sok} is often used technique for obtaining sensitive information. Fraudsters frequently ask victims to verify their account number and PIN for security reasons, or they claim that suspicious activity has been detected on their account, prompting them to provide their most recent transaction information \cite{wood2023analysis}. Another common tactic is to request the victim's social security number and date of birth in the name of updating records.
These tactics take advantage of human psychology by instilling a sense of urgency and leveraging perceived authority, making it difficult for people to identify fraudulent phone calls \cite{malhotra2023detection}. Such fraud has far-reaching consequences, causing not only financial losses but also emotional distress and a significant violation of personal privacy.

Traditional fraud detection methods have primarily relied on rule-based systems and manual monitoring, both of which have limitations in terms of scaling and adaptability to evolving fraud techniques. The introduction of machine learning (ML) and deep learning (DL) has transformed this field, providing dynamic analysis and pattern recognition capabilities that far exceed traditional methods. Neural networks, particularly those based on transformer architectures, have shown remarkable success in detecting anomalies and fraudulent activities by learning from massive datasets of legitimate and fraudulent interactions.

Large Language Models (LLMs) improve these capabilities by understanding and processing natural language with high precision. LLMs, such as OpenAI's GPT-3, can understand context, detect subtle cues in conversations, and predict potential fraudulent behavior using learned patterns. However, their use in real-time fraud detection systems has been limited \cite{weidinger2021ethical}, \cite{jiang2024detecting}, resulting in a missed opportunity to fully realize their potential.
Retrieval-Augmented Generation (RAG) is a cutting-edge technique that combines the power LLMs with the ability to retrieve relevant information from vast databases. RAG’s major benefit is its ability to stay updated with the latest information and policies without needing retraining, ensuring it remains current and effective. 

In our methodology, we use RAG to detect fraud by following policies established by various banks and companies about what employees should and should not ask. This ensures that the LLMs can make informed decisions based on the most recent policies, thereby detecting potential fraud more effectively.
This paper provides a detailed analysis of our RAG-based policy check system using synthetic call recordings. We compare its performance to traditional deep learning models and simpler LLM-based approaches. Furthermore, we discuss the practical application of our system, which includes Automatic Speech Recognition (ASR) and LLM agents, emphasizing its potential to revolutionize real-time fraud detection in phone calls. Our contributions in this paper include:

\begin{enumerate}
    \item \textbf{Adaptive Policy Compliance with RAG}: Unlike traditional LLMs that require retraining to maintain accuracy, our RAG model dynamically adjusts to policy changes in real-time, ensuring continuous compliance and enhanced security without the need for retraining.
    
    \item \textbf{Personalized Organizational Integration}: The RAG system seamlessly integrates with each organization's unique knowledge base, delivering tailored responses that align with specific policies and operational needs, making it a highly personalized solution.
   
    \item \textbf{Enhanced Transparency and Explanatory Capabilities}: Unlike conventional NLP models that merely classify and label, our RAG-enabled LLM provides detailed explanations for detected policy violations, promoting transparency, increasing user awareness, and encouraging better compliance practices.
\end{enumerate}

By addressing these points, our methodology not only enhances fraud detection capabilities but also ensures that the system remains adaptable, transparent, and integrated with organizational policies, providing a comprehensive solution for real-time fraud detection in phone calls.

\section{Literature Review}\label{LR}
The detection of real-time fraud in the telecom industry is a challenging task. Identifying fraudulent schemes is crucial to minimize revenue losses and enhancing customer satisfaction. In recent years, various solutions have been proposed to address this issue. 

\paragraph{Call Fraud Detection using Call data Records}
Wahid et al. \cite{wahid2024nfa} developed an online fraud detection model using a Neural Factorization Autoencoder (NFA) to address the increasing problem of telecom fraud. They combined Neural Factorization Machines (NFM) with an Autoencoder (AE) to model customer calling patterns, adding a memory module to adapt to changing behaviors. Evaluated on a large dataset of call detail records (CDR), authors claim that the proposed model outperformed existing methods. Kumar et al. \cite{kumar2024detecting} focused on detecting fraudulent calls using Supervised Intelligent Learning Algorithms. Authors trained various machine learning models on a dataset (kept anonymous for privacy reasons). Wu et al. \cite{wu2024beyond} proposed LSG-FD, a telecom fraud detection model using latent synergy graph (LSG) learning to detect collaborative and disguised fraudulent behaviors. The model reconstructs synergy-oriented graphs, captures individual calling patterns with an LSTM-based encoder, and addresses imbalance issues with a dual-channel graph learning module and a label-aware sampler. Sayyed et al. \cite{al2024mobile} demonstrated the importance of data visualization in fraud detection, using the PAYSIM dataset to show how early visual analysis helps identify anomalies and validate dataset suitability, ultimately enhancing the accuracy and efficiency of fraud detection systems. Hapase et al. \cite{hapase2024telecommunication} proposed a Fraud Resilient Framework using an Enhanced CNN-based approach for SMS phishing detection in telecommunications. By introducing PCC-PCA for efficient feature extraction and Parameterized ReLU in the CNN architecture, the framework significantly improves accuracy and efficiency in categorizing ham and spam messages compared to existing methods. Ravi et al. \cite{ravi2022wangiri} analyzed Wangiri fraud patterns using both supervised and unsupervised ML methods on a large CDR dataset, finding that classification algorithms excel in detecting specific fraud patterns, though the optimal algorithm may vary depending on the pattern. 

\paragraph{Call Fraud Detection using Call Contents}
Miramirkhani et al. \cite{miramirkhani2016dial} studied technical support scams, using an automated system to identify scam-related phone numbers and domains over eight months. They analyzed scammers' tactics through controlled experiments and proposed countermeasures to help users avoid these scams and assist law enforcement.
Zhao et al. \cite{zhao2018detecting} focused on telecommunication fraud detection by analyzing the content of calls rather than relying on caller numbers. This approach involved using machine learning and Natural Language Processing (NLP) to extract features from collected descriptions of fraud. However, by using basic NLP analysis on call data, the method faces challenges in adapting to real-time changes, potentially limiting its effectiveness in dynamic fraud scenarios. A similar approach was taken by Bajaj et al. \cite{bajaj2019fraud}, who utilized transcribed conversations as features and employed multiple machine learning models to classify fraudulent activities.

\paragraph{Advancements of LLMs in Fraud Detection}
Bakumenko et al. \cite{bakumenko2024advancing} proposed a novel approach to detecting anomalies in financial data by leveraging LLM embeddings. They tested three pre-trained sentence-transformer models to encode non-semantic categorical data from financial records and evaluated five optimized ML models for anomaly detection. Their findings show that LLMs significantly enhance the detection of irregularities in financial journal entries, outperforming baseline models and effectively addressing feature sparsity in the data.
Liming Jiang \cite{jiang2024detecting} explored the use of LLMs for detecting scams like phishing, emphasizing the steps for building a scam detector. Preliminary tests with GPT-3.5 and GPT-4 showed effectiveness in spotting phishing signs, but further refinement and collaboration with cybersecurity experts are needed to address evolving threats. 
In recent years, several studies have highlighted the innovative use of LLMs in the field of fraud detection. These approaches have explored the potential of LLMs to enhance security across various domains, including the detection of outliers in evolving data streams, which are critical in identifying fraudulent activities. Notable works such as \cite{liming2024detecting}, \cite{zhao2024revolutionizing}, \cite{boumber2024llms}, \cite{rinat2023fraud}
and \cite{luz2024enhancing} have demonstrated the effectiveness of LLMs in addressing challenges associated with fraud detection, emphasizing their versatility and adaptability in real-world scenarios.

As advancements in fraud tactics and the increasing ability to bypass CDR technologies pose new challenges, the need for more sophisticated methods has become evident. Traditional CDR-based detection is becoming less effective due to the rise of adversarial AI capabilities, which can easily evade these systems. To address these gaps, our research focuses on analyzing call contents, a relatively underexplored area, for fraud detection. By leveraging LLMs, our methodology seeks to enhance the precision and robustness of fraud detection systems, adapting to the evolving landscape of fraud techniques and providing a more reliable solution in dynamic environments.

\section{Methodology}

\begin{figure}[ht]
\centering
\includegraphics[width=0.67\columnwidth, height=14cm]{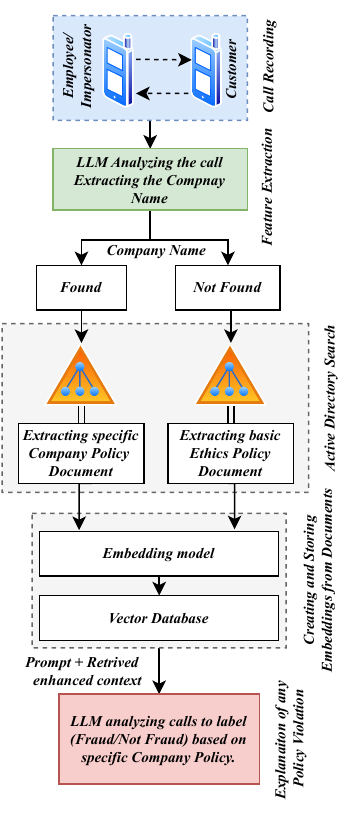}
\caption{Comprehensive Flowchart of the Methodological Framework Employed in the Study}
\label{fig1}
\end{figure}

This section outlines the methodology employed in this research project, covering two key areas. Part 1 focuses on using RAG based LLMs to check policy violations. Part 2 delves into how this approach can be effectively applied in real-time scenarios, while ensuring maximum security and privacy. Figure \ref{fig1} illustrates the core architecture of the proposed method for identifying fraudulent calls. The detailed breakdown of the various components involved in Part 1 of the pipeline is provided in this section, as explained below.

\subsection{Dataset Collection}\label{Dataset}

Privacy and security are critical considerations when working with call records. Call records of some individuals are not public data, and sending them online to a server is a task. So, in our methodology (part 2), we explained how real-time encryption of these audio calls can help send data onto the network with the person's consent while maintaining the necessary privacy. Finding real data (phone calls that were fraudulent or normal) was a challenge for this research. Some researchers have worked with real call data; however, due to legal and ethical concerns, the dataset has not been made publicly available (see Section \ref{LR}). So, in order to test the efficacy of our work, we generated a number of synthetic calls using LLM to simulate real-time scenarios. For our methodology, we used three bank examples: \textbf{Bank1}, \textbf{Bank2}, and \textbf{Bank3}. Each bank provides various services and has its own set of policies. Each bank was given a set of legitimate, intermediate, and fraudulent cases. The LLM was given a prompt using the cases defined in Table \ref{tab:banks}, in order to generate calls with a variety of real-life scenarios based on the defined boundaries. To evaluate our system and compare it to state-of-the-art methods, we generated three sets of synthetic data: 100 calls, 500 calls, and 1000 calls, respectively.

\begin{table*}[!htbp]
\centering
\caption{Legitimate, Fraudulent, and Intermediate activities for different banks}
\begin{tabular}{|l|l|p{10cm}|}
\hline
\textbf{Bank} & \textbf{Type} & \textbf{Description} \\ \hline
\multirow{3}{*}{Bank1} & Legitimate & Confirming contact information, asking for feedback on digital tools. \\ \cline{2-3} 
 & Fraudulent & Requesting full debit card numbers, demanding CVV code and expiry date. \\ \cline{2-3} 
 & Intermediate & Asking for the last 6 digits of account numbers for verification. \\ \hline
\multirow{3}{*}{Bank2} & Legitimate & Verifying identity using the last four digits of the Social Security number, confirming recent transactions. \\ \cline{2-3} 
 & Fraudulent & Requesting immediate transfers of all funds, demanding the customer transfer their balance to an external account. \\ \cline{2-3} 
 & Intermediate & Asking for the last 4 digits of the account number for verification. \\ \hline
\multirow{3}{*}{Bank3} & Legitimate & Asking for feedback on digital tools, participating in surveys about banking services. \\ \cline{2-3} 
 & Fraudulent & Asking customers to download software or remote access tools, demanding the installation of a security application. \\ \cline{2-3} 
 & Intermediate & Requesting the last 6 digits of the customer's account number for verification. \\ \hline
\end{tabular}
\label{tab:banks}
\end{table*}



\subsection{Feature Extraction using LLM}
In our methodology, we use LLMs to analyze and extract key features from call recordings (text). The first step is to extract specific features, such as the employee's name and the company name. Identifying the company name is critical for locating pertinent documents, which are required to explain and label calls as fraudulent or legitimate. Each company has unique policies and privacy concerns, so we must tailor our analysis to the specific company.
After extracting the company name, we address potential issues like spelling errors and variations in the extracted text. To deal with this, we use a cosine similarity check rather than direct matching. 
First, the extracted company name and the names in our database are vectorized using word embeddings. The cosine similarity between two vectors, \( \mathbf{A} \) and \( \mathbf{B} \), is then calculated using the following formula:

\begin{equation}
\text{cosine\_similarity}(\mathbf{A}, \mathbf{B}) = \frac{\mathbf{A} \cdot \mathbf{B}}{\| \mathbf{A} \| \| \mathbf{B} \|}
\label{eq1}
\end{equation}

where \( \mathbf{A} \cdot \mathbf{B} \) is the dot product of the vectors, and \( \| \mathbf{A} \| \) and \( \| \mathbf{B} \| \) are the magnitudes of the vectors. By applying cosine similarity, we effectively match the extracted company name with the closest entry in our database, even if there are minor spelling errors. This ensures accurate document selection for the next steps in our methodology.

\subsection{Retrieval-Augmented Generation (RAG)}

Retrieval-Augmented Generation (RAG) is a potent method that combines text generation and information retrieval capabilities. By incorporating company-specific policy documents, we leverage RAG in our methodology to improve the fraud detection process. In our system, it functions as follows:

\begin{itemize}
    \item \textbf{Company Policy Documents}: As discussed in Section \ref{Dataset}, we do not have real-time policies for each bank/company. So we created a collection of documents for each company outlining their employee policies and guidelines. These documents are critical for providing context-specific explanations and detecting potential fraud according to company standards.
    \item \textbf{Embeddings and Vector Database}: In our system, we use embeddings to represent company policy documents as well as extracted company names from calls. These embeddings are saved in a vector database, allowing us to quickly retrieve relevant documents without having to recalculate embeddings for each query. Storing embeddings in a vector database ensures that our system is efficient and scalable, able to handle large amounts of data and queries.
\end{itemize}

Now vector databases are created. The document embeddings will be fetched directly and passed along with the prompt to the next step, ensuring informed and tailored results.

\subsection{LLM-Based Policy Compliance Check}

With our enriched embeddings ready to use, we move on to the next step, in which we use a prompt to check for policy compliance. The prompt we provide is intended to ensure that the conversation adheres to the company policies. Here's the prompt we used:

\begin{verbatim}
policy_prompt = f"""
You are a policy inspector. Your role is
to ensure that the conversation complies
with the policies provided. Don't use any
external source of information other than
the policies provided to you as context.
Read all the policies carefully, the 
employee may trick the customer into
doing something but their approval means
nothing, stick to the policies mentioned.

Policies:
{context}

Conversation:
{conversation}

Q: Does the conversation break any
policies? If yes, return 'Fraud'. If no,
return 'Normal'.Please provide only the 
label in the first line and start answer
with "Answer:". Provide justification in
the next line starting with the word 
"Justification:".
"""
\end{verbatim}

Using this prompt, we ask the LLM to assign a label indicating whether the conversation is 'Fraud' or 'Normal' based on the policies specified. The LLM also provides a justification for its decision, which aids in detection and raises awareness among users and companies.

\begin{figure}[htbp]
\centering
\includegraphics[width=0.67\columnwidth, height=22cm]{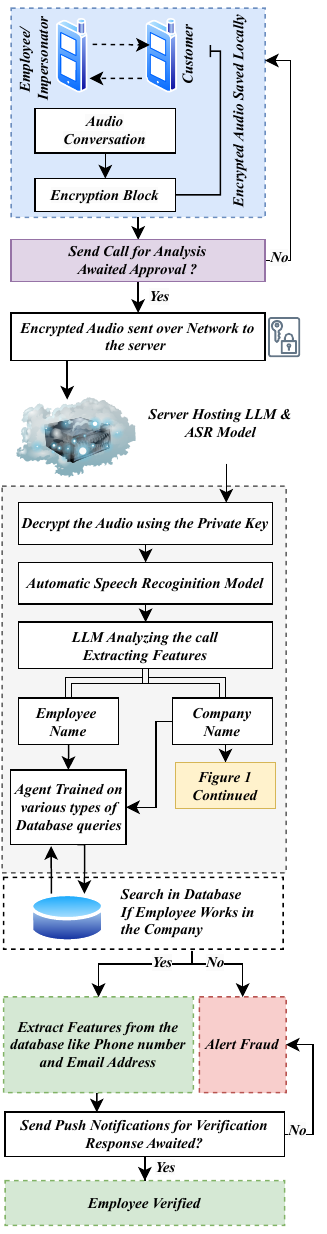}
\caption{Comprehensive Flowchart of the Methodological Framework Employed in the Study}
\label{fig2}
\end{figure}

\section{Real Time Deployment}
This section delves into the practical application of our fraud detection system, emphasizing real-world considerations and challenges. We explain the importance of encryption in protecting sensitive call data, how Automatic Speech Recognition (ASR) is used to transcribe calls, and the specific steps involved in detecting user impersonation. This section aims to provide a thorough overview of how our system works in a realistic setting shown in Figure \ref{fig2}, ensuring both security and accuracy.

\subsection{Encryption Block}

The encryption block is a critical component of our fraud detection system, ensuring that sensitive call data is securely transmitted over the network. Without encryption, data is vulnerable to various attacks, such as Man-in-the-Middle (MitM) attacks \cite{mallik2019man}, where an attacker intercepts and potentially alters the communication between two parties. Encryption ensures that even if the data is intercepted, it cannot be read or tampered with by unauthorized parties, thereby preventing data leaks and enhancing privacy.
In this section, we explain the brief details of the encryption and decryption process and the process of storing encrypted calls locally and obtaining user consent.

\subsubsection{Encryption and Decryption Process}
We use public and private key cryptography to secure call data, ensuring that only authorized parties can access the encrypted information. Our implementation employs AES for data encryption and RSA for securely transmitting the AES key.

\begin{itemize}
    \item \textbf{Encryption Process} : First, we generate a random AES key to encrypt the call recordings. The encrypted call data (\(C\)) and the AES key (\(K\)) are then transmitted securely. The AES key is encrypted using the server's public RSA key (\(K_{enc}\)), ensuring that only the server can decrypt it.
    \item \textbf{Decryption Process} : At the server, the AES key is decrypted using the server's private RSA key. This key is then used to decrypt the call recordings, converting the encrypted data back into its original form.
\end{itemize}

By using AES and RSA together, we ensure that call recordings are securely transmitted and protected against unauthorized access.

\subsubsection{Local Storage and User Consent}
To further protect user privacy, we store encrypted calls locally. During the call, a pop-up message is shown to the user, asking if they want to share the call for analysis. This process allows users to share only the calls they want to analyze, while the rest are removed from the local system, addressing potential storage issues.

\subsection{Automatic Speech Recognition (ASR)}

In our proposed work, we use LLMs, which are specifically designed to understand text and identify patterns in it based on its learning. So, in order to convert the decrypted call on the server to text format, we employ the Automatic Speech Recognition (ASR) model. ASR systems use machine learning models that have been trained on large datasets of audio recordings and transcripts. The primary goal of ASR is to accurately transcribe spoken language in real time, allowing for subsequent analysis and processing.

\subsection{Feature Extraction}

Once the call data is securely decrypted and transcribed into text on the server, the next step involves analyzing the call to extract critical features using a 
LLM. For our application, the primary features of interest are the employee name and company name. These features are essential for further analysis and verification steps. It extracts key information based on predefined prompts and patterns.
Here is an example prompt used by the LLM to extract features:
\begin{verbatim}
Extract the following information from the 
conversation:
- Employee Name
- Company Name

Conversation:
{conversation}

Extracted Information:
- Employee Name: [Name]
- Company Name: [Company]
\end{verbatim}

The extracted feature "company name" is also used in the first part of our methodology (see Section \ref{Dataset}). This includes ensuring compliance with company policies and detecting potential fraud. Because there are no dependent variables, both steps run in parallel following the feature extraction step.




\subsection{User Impersonation Detection}

User impersonation involves an attacker assuming the identity of a legitimate user, typically to deceive others and gain access to privileged information or perform unauthorized actions. This can occur in various ways, such as through social engineering, phishing attacks, or exploiting vulnerabilities in authentication systems. In the context of our system, impersonation can happen when an individual falsely claims to be an employee of a company during a call.

\subsubsection{Detection Steps}
To detect user impersonation, our system uses a two-step verification process. The first step is to check in the database if there is a company name in the database; if so, check to see if the employee works for the company; if that is also true, then proceed to the second verification step, which involves the individual to verify its identity. Both the steps are explained in detail below.

\begin{itemize}
    \item \textbf{First Verification Step: Checking Comapny name and Employee Name} : 
        The first step involves verifying whether the extracted compnay name exists in the public database. Then systems checks if employee name exists in the company's public dataset. This is crucial for ensuring that the individual on the call is indeed an employee of the company.
        \begin{itemize}
            \item \textbf{Cosine Similarity}: Given the limitations of ASR accuracy, we use cosine similarity (explained in eq \ref{eq1}) to match the extracted employee name with the names in the company's public dataset. 
            \item \textbf{Verification}: If the cosine similarity score indicates a strong match, the employee name is considered valid, and we proceed to the next verification step. If not, the call is flagged as potential fraud.
        \end{itemize}

    \item \textbf{Second Verification Step: Sending Push Notifications}: 
        The second step involves verifying the identity of the employee by sending push notifications to their registered phone number and email address. This step ensures that the individual on the call is indeed the person they claim to be.
        
        \begin{itemize}
            \item \textbf{Extracting Contact Information}: From the public dataset, we extract the phone number and email address associated with the verified employee name.
            \item \textbf{Sending Notifications}: Push notifications are sent to the extracted phone number and email address, requesting confirmation of the call. The message may include a unique code or a simple confirmation request.
            \item \textbf{Awaiting Response}: The system awaits a response from the legitimate employee. If the employee confirms the call, their identity is verified, and we rely on the LLM's policy check for further analysis. If the employee does not confirm or rejects the notification, it indicates potential impersonation, and an alert is raised.
        \end{itemize}
\end{itemize}


In this methodology section, we have outlined the comprehensive steps involved in our real-time fraud detection system.
By combining the two components, our methodology provides a robust and scalable solution for real-time fraud detection, ensuring both accuracy and security. This comprehensive approach addresses the various challenges associated with detecting fraudulent activities, making our system a valuable tool for enhancing trust and reliability in communications.

\section{Results}
In this section, we present the results of our fraud detection system evaluated on synthetic call data. As previously discussed (in Section \ref{Dataset}), we created three sets of synthetic data consisting of 100, 500, and 1000 calls respectively. These datasets were used to test the performance of our model with comparison to state of the art methods: A complete Transformer Model (in our case we have used BERT), an untrained (on our data) LLM, and our Retrieval-Augmented Generation (RAG) model.
Table \ref{table:comparison} provides a comprehensive comparison of the models based on various metrics such as Accuracy, Precision, Recall, and F1 Score. The results clearly demonstrate the effectiveness of our RAG model in accurately detecting fraudulent activities.

\begin{table*}[!htbp]
\centering
\caption{Comparison of BERT, Simple LLM Untrained, and RAG models based on different numbers of calls}
\begin{tabular}{|c|c|c|c|c|c|}
\hline
\textbf{Model} & \textbf{Number of Calls} & \textbf{Accuracy} & \textbf{Precision} & \textbf{Recall} & \textbf{F1 Score} \\
\hline
\multirow{3}{*}{BERT} & 100 calls & 0.7800 & 0.7067 & 1.000 & 0.8281 \\
\cline{2-6}
& 500 calls & 0.9360 & 0.9249 & 0.9778 & 0.9506 \\
\cline{2-6}
& 1000 calls & 0.9530 & 0.9479 & 0.9772 & 0.9623 \\
\hline
\multirow{3}{*}{Simple LLM Untrained} & 100 calls & 0.6632 & 0.6875 & 0.2895 & 0.4074 \\
\cline{2-6}
& 500 calls & 0.6605 & 0.6941 & 0.2995 & 0.4184 \\
\cline{2-6}
& 1000 calls & 0.6716 & 0.7013 & 0.2880 & 0.4083 \\
\hline
\multirow{3}{*}{RAG Based LLM} & 100 calls & \textbf{0.9798} & 0.9744 & 0.9744 & 0.9744 \\
\cline{2-6}
& 500 calls & \textbf{0.9776} & 0.9895 & 0.9545 & 0.9717 \\
\cline{2-6}
& 1000 calls & \textbf{0.9766} & 0.9810 & 0.9577 & 0.9692 \\
\hline
\end{tabular}
\label{table:comparison}
\end{table*}

\subsection{Model Performance Analysis}

The performance analysis indicates that the BERT model shows significant improvement with an increasing number of calls, achieving an accuracy of 0.7800 and an F1 score of 0.8281 with 100 calls, and reaching 0.9530 and 0.9623 respectively with 1000 calls. This demonstrates BERT's dependency on larger datasets for higher accuracy and reliability. In contrast, the untrained simple LLM underperforms, achieving an accuracy of 0.6632 and an F1 score of 0.4074 with 100 calls, with only marginal improvements as the dataset size increases, indicating its unsuitability for effective fraud detection without extensive training. However, our RAG model consistently outperforms both BERT and the untrained LLM across all dataset sizes, achieving an impressive accuracy of 0.9798 and an F1 score of 0.9744 with 100 calls, and maintaining high performance with larger datasets. This suggests that the RAG model's ability to integrate relevant document retrieval with text generation allows it to provide more accurate and contextually appropriate responses. This means that while BERT and LLM require substantial training, our proposed approach is scalable and does not require extensive training. 

\subsection{Advantages of LLM-based RAG Model over BERT}

This section discusses the primary advantages of the RAG based LLM model over the BERT model. Three scenarios are used to illustrate this point as showcased in detail in Table \ref{table:rag_vs_bert}.

\begin{table*}[htbp]
\centering
\caption{Comparison of proposed model (RAG based LLM) and BERT responses for different scenarios.}
\begin{tabular}{|p{0.15\textwidth}|p{0.25\textwidth}|p{0.25\textwidth}|p{0.15\textwidth}|}
\hline
\textbf{Scenario} & \textbf{Example Call} & \textbf{LLM-RAG Response} & \textbf{BERT Response} \\
\hline
\textbf{Ease of Adaptation to Policy Changes} & Hello, this is John from Bank1. May I ask you to confirm the last seven digits of your card number for security purposes? & No, the conversation does not break any policies. The request for partial account details (\textbf{last seven digits of the card number}) is allowed as per the Intermediate Verifications policy. & Classified as: Fraud \\
\hline
\textbf{Personalized Answers for each Organisation} & Hello, this is John from Bank2. May I ask you to confirm the last six digits of your card number for security purposes? & Yes, the conversation does break the policy. Specifically, it violates the policy \textbf{"Partial Account Details"} under \textbf{"Intermediate Verifications"}. The policy states that employees may ask for the last 4 digits of the customer's account number or debit card for security verification. However, the conversation asks for the last six digits, which exceeds the allowed limit of four digits. & Classified as: Normal \\
\hline
\textbf{Detailed Reasoning and Explanation} & Hello, this is Sarah from Bank3. We've noticed some unusual activity on your account. Can you confirm the last three transactions you made? Sure, I spent 50 at SuperMart, 30 at GasGo, and 15 at BookWorld. Thank you for confirming. To ensure your security, could you please transfer your funds to a temporary 'safe account' while we investigate?
 & Yes, the conversation breaks the "Software Downloads" and "Security Applications" prohibited request policies. \textbf{The agent is asking the customer to transfer funds to a temporary 'safe account', which implies the use of software or remote access tools.} Furthermore, the agent's language could be interpreted as requesting the customer to install a security application for verification purposes.
 & Classified as: Fraud \\
\hline

\end{tabular}
\label{table:rag_vs_bert}
\end{table*}

\begin{itemize}
    \item \textbf{Scenario 1: Ease of Adaptation to Policy Changes}: Let's assume TrustyBank updates its policy to allow seven digits for verification.
    \begin{itemize}
        \item \textbf{LLM-RAG:} Capable of understanding and incorporating new policies immediately without retraining, ensuring continuous compliance and security.
        \item \textbf{BERT:} Requires retraining on new datasets to include updated policies, which can be time-consuming and resource-intensive, potentially leading to delays.
    \end{itemize}

\item \textbf{Scenario 2: Personalized Responses for Each Organization}: As outlined in Table \ref{tab:banks}, only Secure Bank permits asking for the last four digits, while all others allow the last six digits for verification.
    \begin{itemize}
        \item \textbf{LLM-RAG:} Accurately extracts relevant embeddings and provides personalized responses based on each organization’s policy.
        \item \textbf{BERT:} Struggles to adapt to personalized features due to generalized training, requiring additional focused training or attention mechanisms to incorporate such specific features.
    \end{itemize}

\item \textbf{Scenario 3: Detailed Reasoning and Explanation}

    \begin{itemize}
        \item \textbf{LLM-RAG:} Provides comprehensive explanations for detected policy violations, aiding in transparency and understanding.
        \item \textbf{BERT:} Identifies whether a call is fraud/not but lacks detailed explanations, limiting effectiveness in complex scenarios.
    \end{itemize}
\end{itemize}

The results demonstrate that the RAG model excels in real-time fraud detection, outperforming BERT and an untrained LLM, proving its effectiveness and suitability for real-world scenarios.

\section{Limitations and Future Work}

One of the key limitations of the proposed work is that it has been primarily tested on synthetic data. While synthetic data allows for controlled experimentation and testing of various edge cases, the proposed methodology has not yet been tested on real-world data. The performance of the system in a real-world setting, where data can be far more diverse and less predictable, will need to be carefully evaluated. Additionally, the challenge of LLM hallucination, where the model may generate unnecessary or incorrect outputs, persists. These hallucinations, which may not align with the provided context or policies, remain a critical issue. However, advancements in LLM technology and techniques like prompt engineering have shown promise in mitigating these errors. Future work will focus on deploying this system in real-world environments to validate its effectiveness and improve upon areas such as reducing hallucinations and integrating more secure and accurate verification methods \cite{gressel2024discussion}. Enhancing the Automatic Speech Recognition (ASR) component, especially for multiple languages, is another area that could significantly improve the overall integration and effectiveness of the system in fraud detection. These improvements could make this methodology a groundbreaking approach in integrating AI for security and fraud detection.

\section{Conclusion}
In this study, we have presented a novel real-time fraud detection system utilizing Retrieval Augmented Generation (RAG) technology, addressing the growing concern of AI-enabled scams such as fraudulent phone calls and user impersonations. Our approach leverages RAG’s unique capability to continuously update policies without requiring full model retraining, making it highly adaptable to evolving threats. The system operates on two primary fronts: real-time transcription and policy checking to ensure conversation authenticity, and a robust two-step verification process to confirm caller identity, enhancing both transparency and accountability.
Through rigorous testing using synthetic call recordings, our RAG-based system demonstrated superior performance, achieving an impressive accuracy of 97.98\% and an F1-score of 97.44\%, outperforming existing state-of-the-art methods. These results underscore the effectiveness of our approach and its potential to be deployed in real-world scenarios, providing a strong, flexible, and easily updatable defense against AI-driven fraud.
Future work will focus on expanding the system’s capabilities, including broader applications beyond phone call fraud, and further refining the RAG models to enhance accuracy and efficiency. As AI continues to evolve, our strategies for preventing its misuse must also advance, and this system represents a significant step forward in that ongoing effort.

\bibliographystyle{plain}
\bibliography{main}

\end{document}